\documentstyle[epsfig,twoside,fleqn,espcrc2]{article}

\newcommand{\AmS}{{\protect\the\textfont2
  A\kern-.1667em\lower.5ex\hbox{M}\kern-.125emS}}

\title{Electroweak phase transition by four dimensional simulations\thanks{talk presented by Z. Fodor}}

\author{
F. Csikor
\address{Institute for Theoretical Physics, 
E\"otv\"os University, H-1088 Budapest, Hungary.}
Z. Fodor\thanks{On leave from
Institute for Theoretical Physics,
E\"otv\"os University, Budapest, Hungary.} 
\address{Theory Division, CERN, CH-1211 Geneva 23, Switzerland}
J. Hein 
\address{DESY, Notkestr.\,85, D-22603 Hamburg, Germany}
J. Heitger
\address{Institut f\"ur Theoretische Physik I, Universit\"at M\"unster,
D-48149 M\"unster, Germany}
A. Jaster$^{\ c}$ I. Montvay$^{\ c}$}
       
\begin{document}

\begin{abstract}
The finite temperature phase transition in the SU(2)-Higgs model
at a Higgs boson mass $M_H \simeq 35$ GeV is studied in numerical
simulations on four dimensional lattices with time-like extensions
up to $L_t=5$. $T_c/M_H$ is extrapolated to the 
continuum limit and a comparison with the perturbative prediction
is made. A one-loop calculation to the
coupling anisotropies of the SU(2)-Higgs model on lattices
with asymmetric lattice spacings is presented. Our numerical simulations
show that the above perturbative result is applicable in the
phenomenologically interesting parameter region. 
\end{abstract}

\maketitle

\section{Introduction}

At high temperatures the electroweak symmetry is restored.
Since the baryon violating processes are unsuppressed at high temperatures,
the observed baryon asymmetry of the universe has finally been
determined at the electroweak phase transition. 
Due to the bad infrared features of the theory,
the perturbative approach predicts ${\cal O}(100\%)$ corrections 
and breaks down in the physically interesting region, $M_H>65$ GeV
\cite{pert}.
In recent years lattice Monte Carlo simulations have been used
in three (e.g. \cite{three}) and four dimensions (e.g. \cite{four}) 
to clarify the details of the phase transition. Since the bad
infrared behaviour is connected with the bosonic sector of the
Standard Model, in both cases the fermions are omitted and
the SU(2)-Higgs model is studied. The comparison of the two
approaches gives not only an unambigous description of the
finite temperature electroweak phase transition, but an 
understanding of the non-perturbative features of the reduction step, too. 

In this talk  
two selected new results of the four dimensional formulation are discussed.
In Section 2 the SU(2)-Higgs model
at a Higgs boson mass $M_H \simeq 35$ GeV is studied. 
Our four dimensional, finite temperature lattices have time-like extensions
up to $L_t=5$. We extrapolate $T_c/M_H$ to the
continuum limit and compare the result with the perturbative prediction.
In Section 3 we present  the one-loop calculation to the
coupling anisotropies of the SU(2)-Higgs model on lattices
with asymmetric lattice spacings. We test the results by numerical simulations
and show that the perturbative result is applicable in the
phenomenologically interesting parameter region.

The simulations have been performed on the APE (Alenia Quadrics) computers
at DESY-IFH and on the CRAY-YMP at HLRZ J\"ulich. 

\section{Critical temperature for $M_H \simeq 35$ GeV}

The details of the results presented in this Section can be found in
\cite{35GeV}.

Our studies at Higgs boson masses $\simeq$ 20, 35, 50 GeV showed that
the perturbative predictions are in quite good agreement with 
the lattice results for the jump of the order parameter,
latent heat and interface tension \cite{pert,four,35GeV}.
Unfortunately, the non-perturbative results are obtained
on lattices with temporal extensions $L_t=2,3$ and they 
have typically ${\cal O}(10\%)$ errors, thus
the comparison with continuum perturbative predictions is
quite difficult. 

The situation is much better for the critical temperature. Its value is
$1/L_t$ in lattice units. In order to extract the dimensionless 
$T_c/M_H$ one needs the transition point at a given $L_t$
and the $T=0$ mass of the Higgs boson at that point. 
Both of them can be measured quite precisely.

\begin{figure}[htb]
\epsfig{file=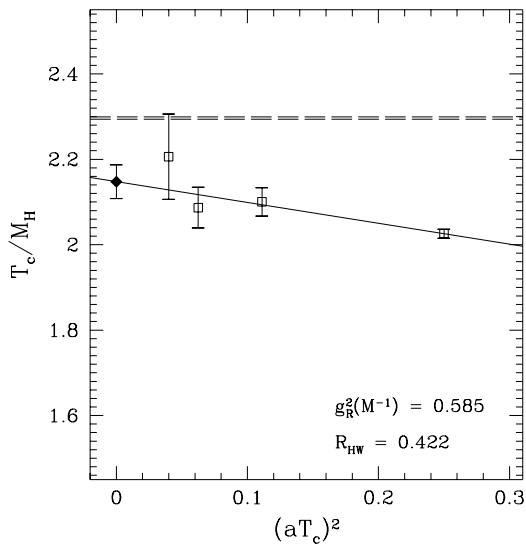,bbllx=40,bblly=220,
bburx=600,bbury=640,width=8.0cm}
\caption{\small Numerical results for $T_c/M_H$ versus $(aT_c)^2=L_t^{-2}$.
The straight line is the extrapolation to the continuum value shown by 
the filled symbol.
The dashed lines are the perturbative predictions at order
$g^3$ (upper) and $g^4$ (lower), respectively. $R_{HW}$ gives
the mass ratio of the Higgs and W boson masses and $g_R^2$
is the renormalized gauge coupling.}
\end{figure}

Fig. 1 shows our results. The non-perturbative 
masses are determined by a careful extrapolation to infinite volumes. 
As it can be seen, the $T_c/M_H$ predictions
of the one-loop and two-loop perturbation theory almost coincide.
The errors of the lattice data are dominated by the uncertainties
in the critical hopping parameter. The extrapolated continuum value 
is $T_c/M_H=2.147(40)$. The value at $L_t=2$ is about 5\% smaller.
This relatively small deviation is better than the expectation
based on lattice perturbation theory. The value of $T_c/M_H$ extrapolated to
the continuum limit differs by about three standard deviations from
the two-loop perturbative result.
This is under the assumption that $L_t=2$ can be included
in the extrapolation, which is supported by the good quality of the
fit ($\chi^2 \simeq 1$).

\section{The SU(2)-Higgs model on asymmetric lattices}

The details of the results presented in this Section can be found in
\cite{pert_ani,non_pert_ani}.

For larger $M_H$ (e.g. $M_H=80$ GeV)
the electroweak phase transition gets weaker, 
the lowest excitations have
masses small compared to the temperature, $T$. From this
feature one expects that a finite temperature simulation
on isotropic lattice would
need several hundred lattice points in the spatial directions
even for $L_t=2$ temporal extension. 
This difficulty can be solved by using asymmetric lattices, i.e. lattices
with different spacings in temporal ($a_t$) and spatial
($a_s$) directions. The asymmetry of the
lattice spacings is characterized by the asymmetry factor $\xi=a_s/a_t$.
The different lattice spacings can be ensured by
different coupling strengths, 
($\beta_s$ and $\beta_t$ for space-space and space-time plaquettes; 
$\kappa_s$ and $\kappa_t$ for space-like and time-like hopping terms)
in the action for different directions (c.f. \cite{pert_ani}). 
The anisotropies $\gamma_\beta^2=\beta_t/\beta_s$ and
$\gamma_\kappa^2=\kappa_t/\kappa_s$ are functions of the asymmetry $\xi$.
On the tree-level the coupling anisotropies are equal to the
lattice spacing asymmetry; however, they receive quantum corrections
in higher orders of the loop-expansion. On the one-loop level one gets
\begin{center}
$\gamma_\beta^2=
\xi^2[1+c_\beta(\xi)g^2+b_\beta(\xi)\lambda]$,\\
$\gamma_\kappa^2=
\xi^2[1+c_\kappa(\xi)g^2+b_\kappa(\xi)\lambda]$, 
\end{center}
where $g$ and $\lambda$ are the bare gauge and scalar coupling, respectively. 

In general, the determination of $\gamma_\beta(\xi)$ and
$\gamma_\kappa(\xi)$
should be done non-perturbatively. This can be achieved by 
tuning the coupling strengths requiring 
that the Higgs- and W-boson correlation lengths in physical units
are the same in the different directions.

\begin{figure} 
\epsfig{file=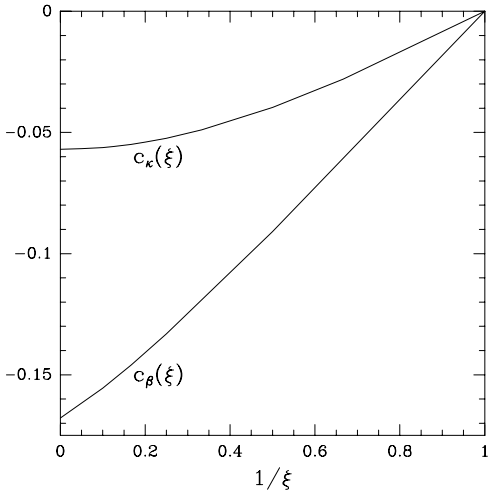,bbllx=20,bblly=220,
bburx=600,bbury=690,width=6.8cm}
\caption{{\small $c_\beta (\xi)$ and $c_\kappa(\xi)$ as functions of
$1/\xi$.}}
\end{figure}

We have used the above idea in perturbation theory.
The one-loop results are summarized on Fig. 2. As it can be seen 
only $c_\beta (\xi)$ and $c_\kappa(\xi)$ are given. The functions
$b_\beta (\xi)$ and $b_\kappa(\xi)$ vanish in this order. Our
calculation is the extension of \cite{karsch} to a gauge-Higgs
model. Omitting graphs connected with the scalar sector,
the result of \cite{karsch} can be reproduced 
(the function $c_\beta (\xi)$ of the present paper corresponds to
$c_\tau(\xi)-c_\sigma(\xi)$ of ref. \cite{karsch}).

It is necessary to check the validity of the perturbative result
by non-perturbative methods. $M_H \simeq 80$ GeV has been chosen, thus
$M_H \simeq M_W$ at zero temperature with typical correlation lengths
of 2-4 in lattice units. We have performed simulations with two different
sets of anisotropies.
First $\gamma_\beta=\gamma_\kappa=4$ (tree-level). Then $\gamma_\beta=3.8$, 
$\gamma_\kappa=4$. As it can be seen on Fig. 3 in the first case 
$\xi_W>\xi_H$ 
($\xi_i=M_i(space-direction)/M_i(time-direction)$, where
$i$ denotes the W or Higgs chanel). In the second case $\xi_W<\xi_H$. 
Rotational invariance is restored if $\xi_W=\xi_H$. Our linear interpolation
is shown by the two solid lines. The ``matching point'' is in complete
agreement with the perturbative prediction (full triangle). In addition
we have simulated at the perturbatively predicted couplings. At this
third point the rotational invariance is restored, thus 
$\xi_W=\xi_H$ within errorbars.  

The non-perturbative features of the theory appear at high temperatures. 
Since the theory is defined at $T=0$, where it is weakly interacting, 
the good agreement between perturbation theory and non-perturbative results
is not surprising. The perturbative predictions open the possibility
to study the phase transition for higher Higgs boson masses and to 
vary the space-like and time-like lattice spacings separately.

Two of us (F.~Cs. and Z.~F.) were partially supported by Hungarian
Science Foundation grant under Contract No.\ OTKA-F1041/3-T016248/7.

\begin{figure} 
\epsfig{file=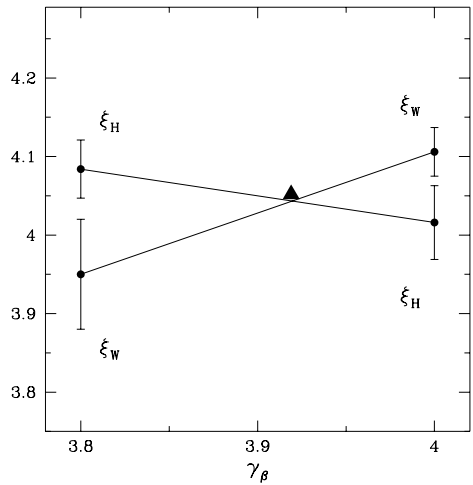,bbllx=20,bblly=220,
bburx=600,bbury=690,width=6.8cm}
\caption{{\small Nonperturbative determination of the anisotropies}}
\end{figure}


\begin{thebibliography}{9}
\bibitem{pert} W. Buchm\"uller, Z. Fodor, T. Helbig, D. Walliser,
Ann. Phys. 234 (1994) 260; 
Z. Fodor, A. Hebecker, Nucl. Phys. B432 (1994) 127; 
W. Buchm\"uller, Z. Fodor, A. Hebecker, Nucl. Phys. B447 (1995) 317.
\bibitem{three}
K. Kajantie et al., Nucl. Phys. B466 (1996) 189; 
J. Kripfganz et al.,  Phys. Lett. B356 (1995) 561; 
F. Karsch, T. Neuhaus, A. Patk\'os, Nucl. Phys. B441 (1995) 629;
O. Philipsen, M. Teper, H. Wittig, Nucl. Phys. B469 (1996) 445.
\bibitem{four}
F. Csikor, et al., Phys. Lett. B334 (1994) 405; B357 (1995) 156; 
Z. Fodor et al., Nucl. Phys. B439 (1995) 147; J. Hein, J. Heitger, 
hep-lat/9605009.
\bibitem{35GeV} 
F. Csikor et al. hep-lat/9601016.
\bibitem{pert_ani}
F. Csikor, Z. Fodor, Phys. Lett. B380 (1996) 113. 
\bibitem{non_pert_ani}
F. Csikor, Z. Fodor, J. Heitger, in preparation.
\bibitem{karsch}
F. Karsch, Nucl. Phys. B205 (1982) 285.
\end{thebibliography}
\end{document}